\documentstyle[preprint,aps]{revtex}
\begin{document}
\def\dg{\dagger}
\def\de{\delta}
\def\beg{\begin{equation}}
\def\en{\end{equation}}
\baselineskip24pt
\pagestyle{headings}
\title{Laughlin Wave Function and One-Dimensional Free Fermions}
\author{Prasanta K. Panigrahi$^*$ and M. Sivakumar$^{**}$}
\address{School of Physics, University of Hyderabad,\\
Hyderabad - 500 134, {\bf India}}

\maketitle
\vskip0.5cm
\begin{abstract}
Making use of the well-known phase space reduction in the lowest
Landau level(LLL), we show that the Laughlin wave function for
the $\nu = {1\over m}$
case can be obtained exactly as a coherent state representation of an
one dimensional $(1D)$ wave function. The $1D$ system consists of $m$ copies
of free fermions associated with each of the $N$ electrons,
confined in a common harmonic well potential. Interestingly, the condition
for this exact correspondence is found to incorporate Jain's parton picture.
We argue that, this correspondence between the free fermions and
quantum Hall effect is due to the mapping of the $1D$ system under
consideration, to the Gaussian unitary ensemble in the random matrix theory.

e-mail addresses: $*$ panisp@uohyd.ernet.in, $**$ mssp@uohyd.ernet.in.
\end{abstract}
\newpage
Laughlin wave function, which describes the incompressible quantum
fluid phase of a two-dimensional ($2D$) interacting electron gas has
enjoyed tremendous success in explaining the observed features of the
fractional quantum Hall effect (FQHE) [1,2]. Jain has given an
interpretation of FQHE in terms of the integral quantum Hall effect (IQHE) of
fractionally charged `partons'[3] and also as IQHE of composite
fermions-- fermions with even number of flux quanta attached [4]. In
recent times there have been attempts [5] to relate quantum Hall effect
with a certain $1D$ model-- the Calegero-Sutherland model
[6]. This is due to the similarities between the structures of the
ground state and the excited states of the two systems. Azuma and Iso,
in particular, have shown that, close relationship exists between the
two wave functions for
a narrow channel quantum Hall system [7] and Iso also has argued for an
universality between the two in the long-wavelength limit [8].
It is interesting to enquire if such a construction
is possible under more general conditions and what is the role of Jain's
picture in this context. In this letter, we obtain an exact
mapping between the Laughlin wave
function for the $\nu={1\over m}$ case and the wave function for $1D$
free fermions, making use of an approach analogous to the
Jain's parton picture.

In arriving at the Laughlin wave function, it is assumed that
electrons are restricted to the lowest Landau level (LLL) due to the
effect of the strong magnetic field and low temperatures. It is well
known [9], and as we shall also show below, that this restriction
converts the configuration space of the electrons to a phase space of
an $1D$ system. Viewed in this way, it is pertinent to ask if the
Laughlin wave function itself can be considered as a coherent state
representation of an $1D$ system. We show that, the ground state wave
function for $m$ non-interacting
partons, associated with each of the N electrons, confined in a common
harmonic well, has as its coherent
state representation the Laughlin wave function when certain
restrictions are imposed. Interestingly, these conditions implement
Jain's parton picture quite elegantly. In contrast to [8], we
do not need any restriction on the sample size.

Consider a particle of charge $e$ and mass $m_0$ in two dimensions $(2D)$
in a
transverse magnetic field $B$ and a potential V(x,y). In the gauge
$\vec A={B\over 2}(-x,y)$
\beg
\L= {1\over 2}m_0  ($ \.x $^2 +$\.y $^2)+ {eB\over 2c}(-y $ \.x $+
$ x\.y$ )-V(x,y).
\en
Since the spectrum is equally spaced with spacings $\hbar
\omega_c=\hbar {eB\over {m_0c}}$, when the potential is weak,
 the restriction to LLL takes place,
when zero mass limit is taken. In this limit, it follows from
the Lagrangian [9]
that
\beg
[x,y]=-il^2_B,
\en
where $l^2_B={\hbar c\over {eB}}$. Thus the phase space is reduced
from the four variables $p_x,p_y,x,y $ to two variables $ X_1,X_2 $ defined
below. These can also
be seen as the guiding centre coordinates of the cyclotron orbit given
by $X_i=x_i$+${l^2_B\over {\hbar}}\epsilon_{ij}\pi_j$,
where ${\pi_j= p_j -{e\over c} A_j }$, satisfying the same relation as in
(2). Here, $x_1$ and $ x_2$
are the coordinates (x,y)  of the electrons before LLL
restriction is made. When restricted to the LLL, the coordinates of
the electron in $2D$ are identified with that of the guiding centre
coordinates.  The two coordinates thus
behave like canonically conjugate variables of an $1D$ system. The
combination ${(X_1-(+) i X_2) ={\sqrt 2}b(b^\dagger)}$ obeying the
oscillator algebra, connects the degenerate angular momentum states.
Thus the wave function of LLL state is given by $\Psi(z)=<\Psi|z>$ where
$z=x+iy$ is the eigenvalue of $X_1+iX_2$ on the state $|z>$. $|z>$ is
the coherent state associated with the angular momentum algebra. The
coordinate space represention of the coherent state, $<q|z>$, where
$|q>$  is the $1D$ coordinate basis, follows easily.
Taking the inner product of $<q|$, with the defining coherent state relation
and using a suitable represention for $X_1,X_2$ interms of $1D$ variables,
the following result follows.
\beg
 <q|z> ={1\over {\sqrt (l_B \sqrt \pi)}}\exp{-{1\over 2 {l_B}^2} }({\bar z z
+ z^2+ q^2 -2 {\sqrt 2} qz}).
\en
The Laughlin wave function for $\nu ={1\over m}$ is given by
\beg
<\psi |z_i>={\prod_{i \ge j}}(z_i- z_j)^m \exp - \sum_i {\bar z_i z_i}
\en
where  we have expanded the state $<\psi|$  in the coherent state
basis. We wish to identify the $1D$ system whose coherent state represention
gives  exactly (4).

Consider the ground state wave function of $ m $ non-interacting fermions
associated with each of the N particles, confined in a harmonic well of
frequency $ \omega $ given by the cyclotron frequency. These $m$ particles are
referred to as `partons' following Jain and the reason for that will
be clear later.
\beg
< \psi|q> \equiv \psi({q_1}^{(1)}..{q_1}^{(m)}\cdots{q_N}^{(1)}..{q_N}^{(m)}) =
{\prod_{ i \ge j ,a}}
  ( {q_i^{(a)} -q_j^{(a)}})
\exp- ({m_0  \omega^2\over 2\hbar}\sum_{i,a}q^2{{_i}^{(a)}})
\en

This wave function, which is the ground state wave function of free
fermions in a common harmonic well, is anti- symmetric in particle index
$i,j$ and symmetric in the` parton' index, $a$. The number of partons
$ m $,has to be odd for it to describe electrons in LLL in $2D$.
We show that the coherent state representation of (5) is (4) when the
`partons ' are constrained to be the same in the phase space, for each
of the N particles and their charges  chosen to be $1/m$ of electrons.

  To find the coherent space representation of (5), we need,
\beg
<\psi|z> = \int <\psi|q><q|z> dq.
\en
 The overlap between the
coherent state and the coordinate space is given by

\beg
\ <q|z>  \equiv  \prod _{i,a}<q_i^{(a)}|z_i^{(a)}>.
\en
This is a generalization of (3) from the one particle case to the many
particle
case. This expression in conjunction with (5) gives the coherent space
representation of a given wave function.
${z _i}^a$ has now  the meaning of coordinates of `partons' in LLL
in $2D$. The chirality associated with the Laughlin wave function, due to the
presence of the magnetic field, enters in (7 ): if the direction of the
magnetic field is reversed, then  $z \rightarrow \bar z$, as $|z>$ is the
coherent
state associated with the angular momentum lowering operator.
Now  $ {z_i}^a$ are chosen to be the same for all $ (a)$ for a given $i$:
\beg
\ z_i^{(a)} = {z_i}.
\en
Only with this restriction the Jastrow factor of the coherent state
 representation matches with that of the Laughlin wave function.
This choice has the physical meaning of constraining the coordinates of
the $m$ `partons' to be the same in the $2D$ coordinate space of the LLL
system. This restriction is anologous to Jain's picture of treating
electrons as composed of $m$ partons. This is the reason for the term
parton used in this letter.

 Also as in the Jain's picture, each parton having coordinate $z_i$ in
LLL is taken to have charge $1/m$ of electrons. This is needed, as we
shall see, to obtain Laughlin wave function with the correct Gaussian width.

    Evaluating (6), using (8) in (7), requires the following result:

\begin{eqnarray}
\int\, \prod_i dq_i \prod_{i<j} (q_i-q_j)\exp{-\sum_i
(q_i^2-\sqrt2 q_iz_i)}\,\,\nonumber\\
=const \prod _{i>j} ({d\over {d z_i}} -{d\over {d z_j}})
\exp \sum_i( z_i^2)
= const\prod _{i>j}(z_i-z_j) \exp { \sum}_i (z_i^2).
\end{eqnarray}

Using (9), to evaluate (7) by integrating over each  $1D$ coordinate
of the partons for all of the N particles, we get
\beg
<\psi|z>=const\prod_{i>j}(z_i - z_j)^m \exp-{m\sum_i|z_i|^2\over {2l_B^2}}.
\en
By choosing the charge of each parton to be ${1\over m}$ of that of
the electron's charge, Laughlin wave function results.

We have thus shown, using the LLL restriction, that the Laughlin wave
function is the holomorphic reprsentation of an $ 1D$ system of free
fermions in a harmonic well. The fact that it is the non-interacting
fermions, which are  related to FQHE wave functions, can  possibly be
understood as follows. It is
well known
that the probability distribution of free fermions in harmonic
confinement in $1D$, is isomorphic to the probability distribution of the
eigenvalues in a Gaussian unitary ensemble in the Random Matrix theory
[10]. The latter corresponds to time-reversal non-invariant Hamiltonian
systems. In recent times [11], an intriguing connection has been
established between static and dynamic correlations of eigenvalues in Random
Matrix theory and particle coordinates in Calegero-Sutherland
model for certain values of the coupling constants, which includes
free fermions. Viewed in this light, therefore, it is not surprising that,
the eigenvalue distribution in {\it time reversal non- invariant } ensembles
correspond to the probability distribution of coordinates of
electrons in a {\it magnetic field}.

   Our construction of Laughlin wavefunction from $1D$ fermionic theory
is different from that of the consrtuction in [7] : in our construction
$1D$ fermions are free
fermions, but having $m$ of them associated with each of the $2D$
electrons; it is an exact correspondence, valid independent of the
sample size. The restrictions needed to obtain it and the value
of the parton's  charge, correspond to the Jain's picture. This
relation of Laughlin wave function and $1D$ systems is possibly related,
in general, to the edge states of Quantum Hall effect and specifically
to [12], where FQHE is related to $1D$ free fermions.

 Extension to other filling factors involves, in Jain's picture,
filling up of higher Landau levels. Hence, their identification to $1D$ systems
through  the phase space picture is not obvious. It should be
interesting to find such  an extension for other filling factors. This
correspondence may also be useful to study the symmetry aspects of FQHE,
like W$_\infty $ symmetry [13] since the $1D$ system we have is known
to have such a symmetry [14]. This can also, possibly, offer a better
calculational procedure to compute expectation values in quantum Hall
states by converting them to $1D$ problems.

Acknowledgement:
We thank S. Chaturvedi and V. Srinivasan for useful discussions.
MS acknowledges R. Shankar and M.V.N. Murthy for an illuminating discussion.

\noindent{\bf References }
\begin{enumerate}

\item  R.B. Laughlin, in {\it The Quantum Hall Effect}, edited by
	R. Prange and S. M. Girvin (Springer-Verlag, New York, 1989),
	and references therein.
\item   T. Chakraborthy and P. Pietilainen,{\it The Fractional Quantum
	Hall Effect} (Springer-Verlag, Berlin, 1988).

\item   J.K.Jain,  Phys. Rev.{\bf B 40 },8079 (1989);\\
       X.G.Wen,Int.J.Mod.Phys.{\it B6},1711(1992).
\item  J.K.Jain, Advances in Physics {\bf 41}, 105 (1992).

\item  N.Kawakami, Phys.Rev.Lett.{\bf71},937(1991),
	J.Phys.Soc.Jpn.{\bf62},2270,2419(1993).

\item   F. Calegero, J. Math. Phys.{\bf10}, 2197 (1969);\\
	B. Sutherland, J. Math. Phys.{\bf12}, 246 (1971).

\item H.Azuma and S.Iso Phys .Lett{\bf B 331}, 107 (1994).
 \item  S.Iso, University of Tokyo Preprint, UT-676.
\item  S. M. Girvin and T. Jach, Phys. Rev. {\bf B 29}, 5617 (1984);\\
	C. Itzykson, in {\it Quantum Field Theory and Quantum Statistics},
	edited by  A. Hilger ( University Press, Briston, 1986);\\
	T. H. Hansson, J. M. Leinaas and J. Myrheim, Nucl. Phys. {\bf B 384},
	559 (1992);\\
	S. Iso, D. Karabali and B. Sakita, Nucl. Phys. {\bf B 385}, 700
	(1992);\\
	G. V. Dunne and R. Jackiw, Nucl. Phys. B (Proc. Suppl.)
	{\bf 33C}, 114 (1993).
\item   M.L.Mehta ,{\it Random Matrices}( Academic Press, New York,1991).
\item   B.D.Simons,P.A.Lee and B. Altshuler, Phys.Rev.Lett.{\bf 70},4122,
	{\bf72},64 (1994).
\item  A.P.Balachandran, L.Chandar, and B.Sathiapalan, Syracuse
	University  Preprint, SU-4240-578.
\item  S. Iso, D. Karabali and B. Sakita, Phys. Lett.{\bf 296}, 143 (1994);\\
	A. Capelli, C. A. Trugenberger and G.R. Gemba, Nucl.Phys.{\bf B396},465
	 (1993).
\item S.Das, A.Dhar, G.Mandal and S.R.Wadia, Int. J. Mod. Phys.{\bf A7},
	5165 (1992).

\end{enumerate}

\end{document}